\documentclass[lettersize,journal]{IEEEtran}
\usepackage{amsmath,amsfonts}
\usepackage{algorithmic}
\usepackage{algorithm}
\usepackage{array}
\usepackage{textcomp}
\usepackage{stfloats}
\usepackage{url}
\usepackage{verbatim}
\usepackage{graphicx}
\usepackage{cite}
\usepackage{makecell}
\usepackage{color}
\usepackage{subcaption}
\usepackage[none]{hyphenat}
\usepackage[table]{xcolor}
\usepackage{colortbl}
\usepackage{flushend}

\usepackage[top=18mm, left=12mm,right=12mm, bottom=17mm]{geometry}

\newif\ifrev
\revtrue 
\ifrev

\else

\fi

\begin{document}
\bstctlcite{IEEEexample:BSTcontrol}
\title{Vertical Assessment of RF-EMF Exposure in a Building Adjacent to a Multi-Operator Shared Base Station}

\author{Ricardo~Q.~de~F.~H.~Silva,~Marcio~E.~C.~Rodrigues,~Fred~S.~R.~Pinheiro, Gutembergue~S.~da~Silva,~Halysson~B.~Mendonça~and~Vicente~A.~de~Sousa~Jr.       
\thanks{Halysson B. Mendonça (ORCID: 0000-0002-9411-2809) is with the Brazilian National Telecommunication Agency (ANATEL), Brazil (e-mail: halysson@anatel.gov.br). Ricardo Q. de F. H. Silva (ORCID: 0000-0003-0861-4341), Marcio E. C. Rodrigues (ORCID: 0000-0002-8399-3241), Fred S. R. Pinheiro (ORCID: 0000-0002-4442-8784), Gutembergue S. da Silva (ORCID: 0000-0001-5951-4805), and Vicente A. de Sousa Jr. (ORCID: 0000-0003-2859-6136) are with the Federal University of Rio Grande do Norte, Brazil (e-mails: ricardo.queiroz.105@ufrn.edu.br, \{marcio.rodrigues, fred.rossiter, gutembergue.soares, vicente.sousa\}@ufrn.br).}  
\thanks{This study was financed in part by Coordena\c{c}\~{a}o de Aperfei\c{c}oamento de Pessoal de N\'{i}vel Superior (CAPES) - Brazil - Finance Code 001. In compliance with CNPq Ordinance No. 2664/2026, the authors attest that generative artificial intelligence was used solely for linguistic editing and text refinement, with no usage in the study’s conceptualization, analysis, or scientific interpretation.} 
\thanks{The authors thank ANATEL for the partnership, with special acknowledgment to its operational unit in Natal.}
\thanks{Submission: 2025-11-16, First decision: 2026-02-10, Acceptance: 2026-03-20, Publication: 2026-03-27.}
\thanks{Digital Object Identifier: 10.14209/jcis.2026.8}
}

\setcounter{page}{77}
\markboth{JOURNAL OF COMMUNICATION AND INFORMATION SYSTEMS, VOL. 41, No.1, 2026.}{Silva \MakeLowercase{\textit{et al.}}: Vertical Assessment of RF-EMF Exposure in a Building Adjacent to a Multi-Operator Shared Base Station}

\maketitle

\begin{abstract}
This paper extends a previously published proposed approach for estimating worst-case exposure scenarios in buildings using publicly available Base Station (BS) parameters, applying it to a critical case involving a multi-operator shared site. Radiofrequency Electromagnetic Fields (RF-EMF) exposure was assessed on every floor of a 20-story building facing the site, revealing significantly higher levels near the center of the antennas’ estimated vertical zone of interest, with a peak of 31.86 V/m, 23.95 times higher than the street-level peak value. However, due to uncertainties in the input parameters, measurements across all locations within the zone of interest are required to accurately identify the floor of maximum exposure. The results demonstrate the usefulness of the proposed estimation approach and highlight the need for improved methods that consider BS antenna configurations and current technical information to ensure accurate assessments of exposure to RF-EMF.

\end{abstract}


\section{Introduction}
\label{sec:introduction}

The assessment of Radiofrequency Electromagnetic Field (RF-EMF) exposure in urban areas has been the subject of numerous studies worldwide~\cite{systematic_review_1}. Particular emphasis has been placed on exposure inside buildings near Base Station (BS) antennas, since the apartments in these buildings have a higher exposure potential.

Several studies have approached these scenarios, either as the primary topic of investigation or as part of more general investigations. Some studies rely on scenario modeling~\cite{prediction_model_2012, amsterdam_2014_1}, seeking predictive models that can estimate exposure levels from antenna configuration, urban morphology, and spatial relationships. Others adopt in situ measurement campaigns~\cite{university_of_macau, romenia_2010, lithuania_2012, turquia_2016, turquia_2018, sono_2021, trabalho_andre, france_2021, suecia_2022, drone_2023, france_2023, grecia_2024}, providing empirical characterization of exposure levels in buildings under real deployment conditions. 

In the second group, the works that employ technical information from BS antennas to determine assessment points in buildings are limited to measurements in a single building, with full knowledge of the exact antenna configuration and installation parameters~\cite{prediction_model_2012, lithuania_2012}. In studies where researchers assessed more than one building~\cite{sono_2021, france_2021}, they used the technical data solely to identify which buildings fall within the exposure zone of antenna emissions, without estimating whether specific floors may be more susceptible to worst-case exposure scenarios.

An author's previous study~\cite{STOTEN_2025} proposed a new approach for assessing RF-EMF exposure in buildings directly exposed to emissions from BS antennas. This method uses BS antenna configuration parameters to select buildings that may be subject to higher exposure levels and to estimate which floors are likely to experience the worst-case exposure. The selection procedure, based on publicly available data on BS configuration and geographical location, prioritizes assessments on buildings that are directly exposed. This allows large-scale evaluations, covering an entire city, since measurements are taken only in buildings that meet the approach's criteria. In 2024, we applied the proposed approach in Natal, the capital of the Brazilian state of Rio Grande do Norte. Measurements carried out in four of the city's 22 target buildings revealed average electric field strengths of up to 13.14~V/m and peak values of 22.79~V/m~\cite{STOTEN_2025}. 

For this contribution, we reapplied the proposed approach to the same city in 2025 to track changes in BS deployments. Measurements conducted in one of the selected buildings, not included in the 2024 measurement campaign, drew particular attention for exhibiting a peak electric field strength higher than the maximum reported in the literature for a similar scenario, as presented in~\cite{suecia_2022}. The present study also aims to validate the methodology presented in \cite{STOTEN_2025} through a measurement campaign not focused on the specific points originally defined by the methodology, but rather on adjacent locations that allow continuous vertical assessment under comparable exposure conditions. This novel study extends the results of~\cite{STOTEN_2025} by reporting findings from the unpublished 2025 measurement campaign covering all floors of a building directly exposed to BS antenna emissions. It provides additional empirical validation of the proposed methodology, offers full-building exposure characterization in a high-exposure scenario, and demonstrates the vertical amplification effects in the exposure distribution.

The remaining sections of this work are organized as follows. Section~\ref{sec:scenario} describes the evaluated scenario, while Section~\ref{sec:procedures} presents our measuring methodology. Section~\ref{sec:results} presents and discusses the measurement results. Finally, Section~\ref{sec:conclusions} provides our final remarks with suggestions of future research directions.

\vspace{-5pt}
\section{Evaluated Scenario}
\label{sec:scenario}

We carried out measurements in one of the 34 buildings, identified by reapplying the approach proposed in~\cite{STOTEN_2025}. We identified this building as the highest criticality due to its proximity and the number of nearby telecommunications installations. The target building is located across the street from a two-BS shared site, whose information is given in the Tab.~\ref{tab:BS_information}. We calculated the distance between the target building and the target BSs using the geodesic distance between each object's coordinates. Using the BSs' horizontal zone of interest diagram, we determined which azimuth emissions were directed toward the target building. Using data from ANATEL's Mosaico platform~\cite{anatelsite}, we identify the mobile communication technologies employed in the emissions and calculate the total power of potential transmitters related to the incident azimuth.

\begin{table}[ht]
    \centering
    \caption{Target BSs information.}
    \label{tab:BS_information}
    \begin{tabular*}{\columnwidth}{@{\extracolsep\fill}cc@{\extracolsep\fill}}
        \hline
        \textbf{Parameter} & \textbf{Value}\\
        \hline
        Distance from the Target Building & 44.90~m  \\
        Incident Azimuth & 
        120$^\circ$~\cite{anatelsite} \\
        Azimuth's Total Transmitter Power & 585.6 W~\cite{anatelsite} \\
        Azimuth's Network Technologies~ & GSM, WCDMA, LTE and NR~\cite{anatelsite}  \\
        \hline
    \end{tabular*}
\end{table}

\begin{figure*}[!hb]
    \vspace{-5pt}
    \centering
        \includegraphics[width=0.85\linewidth]{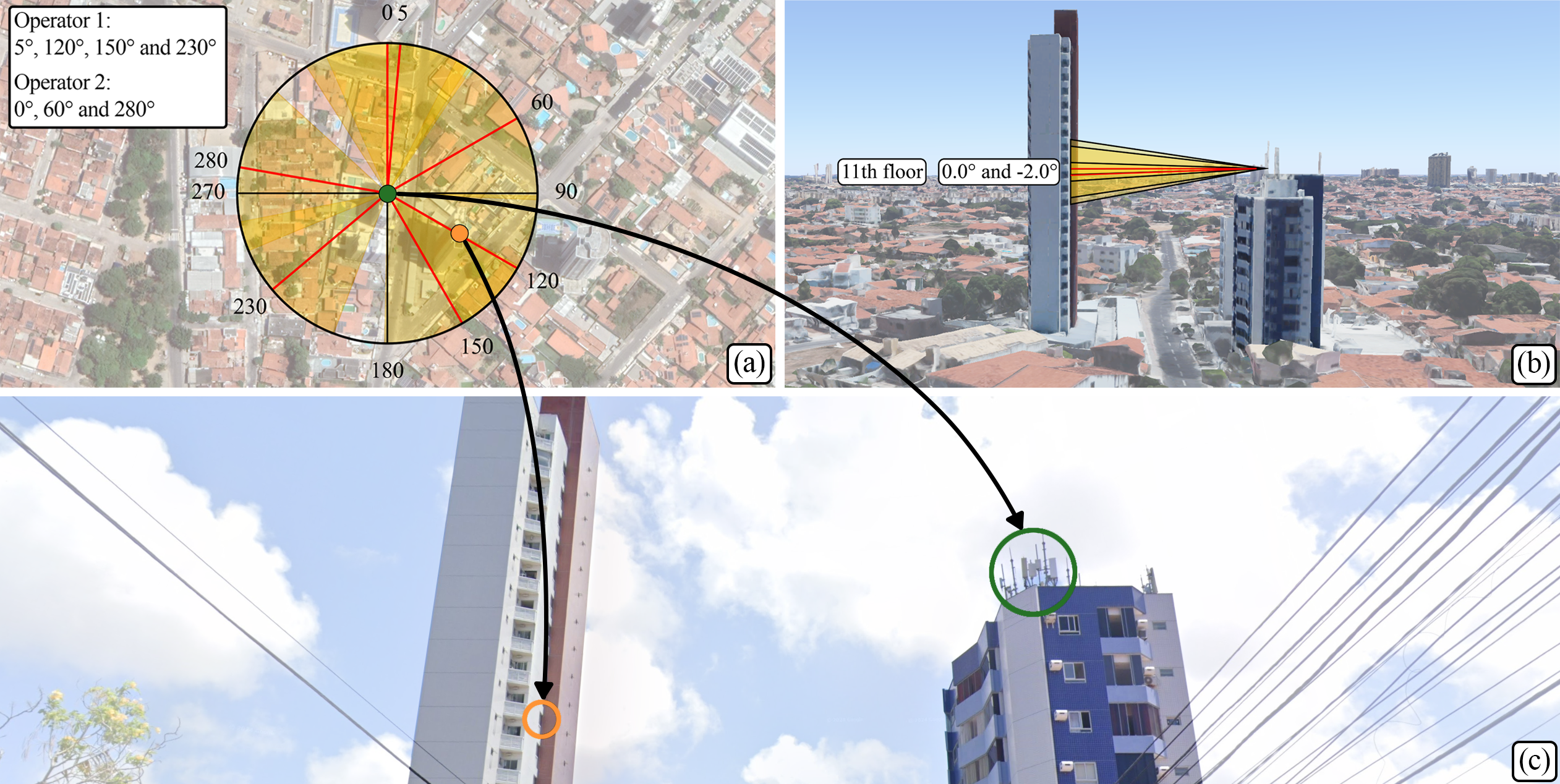}
    \caption{Target building scenario. (a) Target BSs horizontal zone of interest. (b) Target BSs vertical zone of interest. (c) 11\textsuperscript{th} floor (highlighted in orange) and target BSs antennas (highlighted in green).}
    \label{fig:figurapiorcaso}
\end{figure*}

According to~\cite{anatelsite}, the antennas of azimuth 120° radiate at two distinct elevations (-2° and 0°). Overlaying the vertical zone-of-interest diagram on the 3D map image of the site reveals that the center of both emissions falls on the 11\textsuperscript{th} floor of the target building, as shown in Fig.~\ref{fig:figurapiorcaso}b. The target building is also located at the center of the horizontal zone of interest of its respective target BSs, as shown in Fig.~\ref{fig:figurapiorcaso}a. Fig.~\ref{fig:figurapiorcaso}c depicts the position of the 11th floor in relation to the target BSs antennas.

\section{Measurement Procedure}
\label{sec:procedures}

The target building's architectural design allowed measurements in a technical area, without entering the apartment units. This part of the building has free access to the façade, allowing us to conduct one-minute time-averaged measurements on all levels. At each measurement location, the nearby walls are built of brick and covered with ceramic tiles on the outside. Based on the results of this assessment, we identified the floors with the highest electric field strength. Then, we took 30-minute measurements on the three floors with the highest electric field levels. Tab.~\ref{tab:campaigns} displays measurement data for these campaigns.

\begin{table}[!ht]
    \centering
    \caption{Measurement Information.}
    \label{tab:campaigns}
    \begin{tabular*}{\columnwidth}{@{\extracolsep\fill}ccc@{\extracolsep\fill}}
        \hline
        \textbf{Parameter} & 
        \textbf{One-minute Campaign} & 
        \textbf{30-minute Campaign} \\
        \hline
        Date & 02/21/2025 & 02/25/2025 \\
        Start Time & 3:10 PM & 2:40 PM \\
        Finish Time & 4:20 PM & 4:54 PM \\
        Region Temperature & 29~$^\circ$C & 31~$^\circ$C \\
        \hline
    \end{tabular*}
\end{table}

We utilized the Narda NBM-520 broadband field meter~\cite{narda} and the EF 0691 electric field probe~\cite{sonda} to conduct the measurements. This setup enabled the measurement of Root Mean Square (RMS) values of electric field intensity at frequencies ranging from 100~kHz to 6~GHz. We also used a wooden tripod to stabilize the equipment at the measurement points and minimize disturbances during the measurements.

\begin{figure}[!ht]
    \centering   
    \includegraphics[width=0.85\linewidth, trim={0 0 0 2.2cm}, clip]{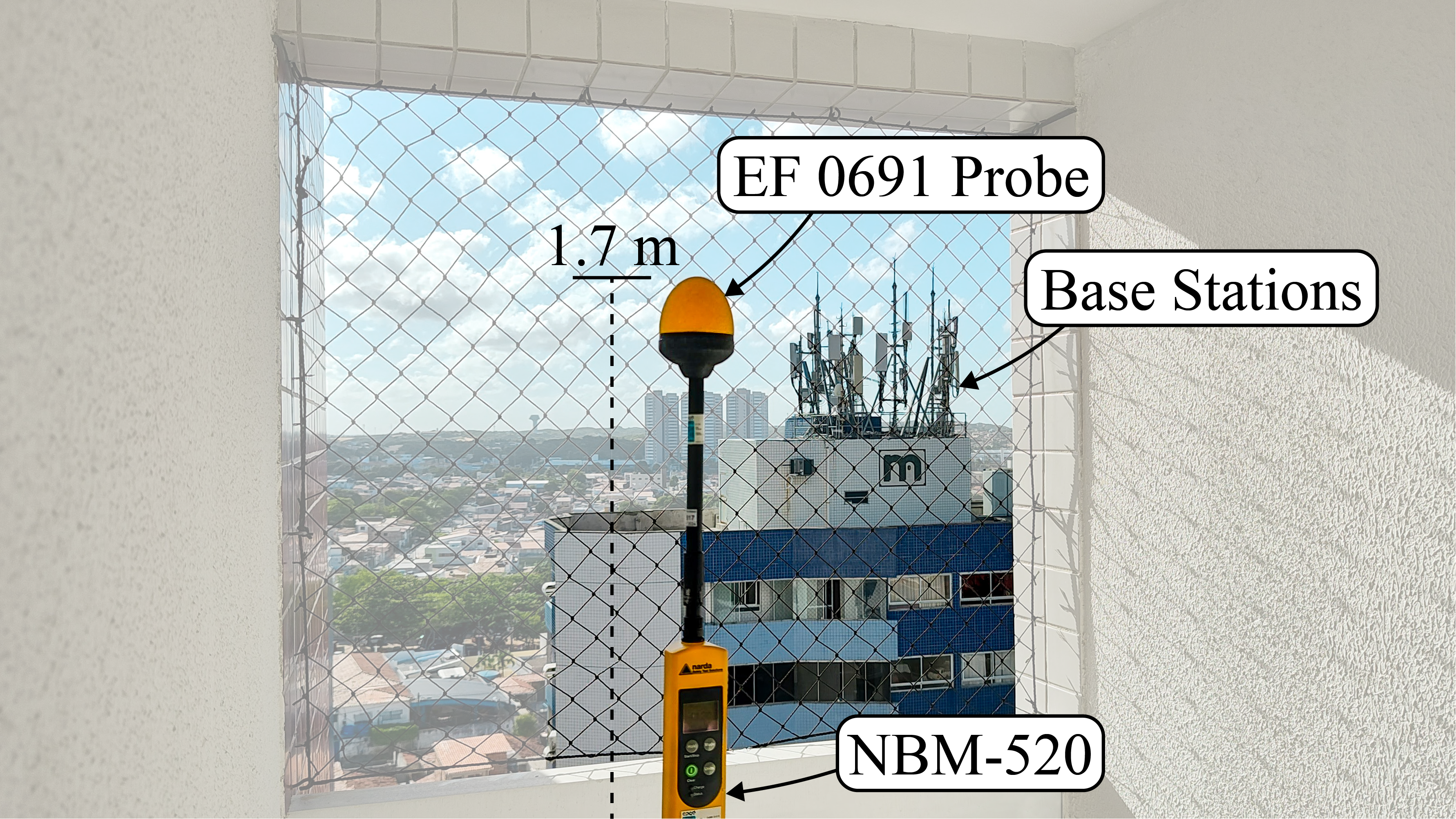}
    \caption{Measurement setup positioned on the 11\textsuperscript{th} floor of the target building.}
    \label{fig:andar11}
    \vspace{-7pt}
\end{figure}

We positioned the measurement setup so that the probe remained at least 2~m away from reflective or absorptive surfaces of electromagnetic waves and from people. In addition, we set the probe height to 1.7~m, as specified in~\cite{ato_17865}.  Fig.~\ref{fig:andar11} shows the measurement setup positioned on the 11\textsuperscript{th}~floor, highlighted in orange in Fig.~\ref{fig:figurapiorcaso}c, during the measurement. Notice that the target BSs antennas are at a height equivalent to the target floor identified by the proposed approach.

To comply with the non-disturbance condition of the measurements, the measurement team kept their mobile phones in airplane mode throughout the entire measurement procedure.

Tab.~\ref{tab:limitesanatel} displays the limits used by Brazilian legislation~\cite{ato_17865}. According to~\cite{ato_17865}, using broadband measurement equipment ensures compliance with the simultaneous exposure limit to multiple radiofrequency fields by not exceeding the most rigorous exposure limit within the assessed frequency range. As a result, the outcomes of this study were established using a limit of 27.70~V/m, which corresponds to the 88-108~MHz broadcasting services operating in Natal.

\begin{table}[!ht]
    \centering
    \caption{General population exposure limits used by ANATEL~\cite{ato_17865}, following~\cite{ICNIRP-2020}.}
    \label{tab:limitesanatel}

	\begin{tabular*}{\columnwidth}{@{\extracolsep\fill}cccc@{\extracolsep\fill}}
	\hline 
	\multicolumn{1}{l}{\textbf{\makecell{Frequency \\ Range}}} & 
	{\textbf{\makecell{Electric Field \\  Intensity,\\ E (V/m)}}} & 
	{\textbf{\makecell{Magnetic Field \\Intensity,\\ H (A/m)}}} & 
	{\textbf{\makecell{Incident Power \\Density, \\S$_{inc}$ (W/m$^2$)}}} \\
    \hline
    8.3~kHz to 100~kHz & 83 & 21 & NA* \\
	0.1~MHz to 30~MHz & $300/f^{0.7}$ & $2.2/f$ & NA* \\ 
    30~MHz to 400~MHz & 27.70 & 0.073 & 2 \\ 
	400~MHz to 2000~MHz & $1.375f^{1/2}$ & $0.0037f^{1/2}$ & $f/200$ \\
	2~GHz to 300~GHz & NA* & NA* & 10 \\ 
    \hline
	\end{tabular*}

    \vspace{1mm}
    {\raggedright *``NA'' signifies ``Not Applicable''. \par}
\end{table}

We also determined the exposure limits for the frequency bands used by the operators of the target BSs, which are available on ANATEL's Mosaico platform~\cite{anatelsite}. Applying the equation from Tab.~\ref{tab:limitesanatel}, we found that the most restrictive limit for these bands is 38.35~V/m, referred to the 778~MHz frequency band used for LTE in the target BSs, as detailed in Tab.~\ref{tab:bandas}. For comparison and informational purposes, we also used this limit in our study findings.

\begin{table}[!ht]
    \centering
    \caption{Limits for downlink frequencies used in the target BS~\cite{anatelsite}.}
    \label{tab:bandas}
	\begin{tabular*}{\columnwidth}{@{\extracolsep\fill}cccc@{\extracolsep\fill}}
	\hline 
	\multicolumn{1}{l}{\textbf{\makecell{Frequency (MHz)}}} & 
	\textbf{\makecell{3GPP \\ Band}} & 
	\textbf{\makecell{Network \\ Technology(ies)}}  & 
	\textbf{\makecell{Limit (V/m)~\cite{ato_17865}}} \\
	\hline
    778.0 & n28 & LTE & 38.35 \\
    874.5 & n5 & WCDMA & 40.66 \\
    885.0 & n5 & WCDMA & 40.90 \\
    1830.0 & n3 & GSM & 58.82 \\
    1855.0 & n3 & GSM and LTE & 59.22 \\
    1875.0 & n3 & GSM and LTE & 59.54 \\
	2117.5 & n1 & WCDMA & NA* \\
    2130.0 & n1 & LTE & NA* \\
    2140.0 & n1 & WCDMA & NA* \\
    2640.0 & n7 & LTE & NA* \\
    2655.0 & n7 & LTE & NA* \\
    3350.0 & n78 & NR & NA* \\
    3550.0 & n78 & NR & NA* \\
    \hline
	\end{tabular*}

    \vspace{1mm}
    {\raggedright *``NA'' signifies ``Not Applicable''. \par}
\end{table}

To match the physical quantity used in ANATEL's RF-EMF evaluations~\cite{anatel_mapa}, we measured electric field intensity (V/m). In the post-processing, we also evaluated the Exposure Ratio (ER), defined in Equation~\ref{eqn:exposure_ratio}~\cite{k100}.

\begin{equation}
\label{eqn:exposure_ratio}
ER = \left(\frac{E_{\text{rms}}}{E_{\text{lim}}}\right)^2
\end{equation}

In Equation~\ref{eqn:exposure_ratio}, $E_{\text{rms}}$ represents the RMS value of the measured electric field (in V/m), and $E_{\text{lim}}$ denotes the reference limit for public exposure established by regulatory guidelines.

\begin{figure}[!ht]
\centering
\includegraphics[width=0.9\linewidth]{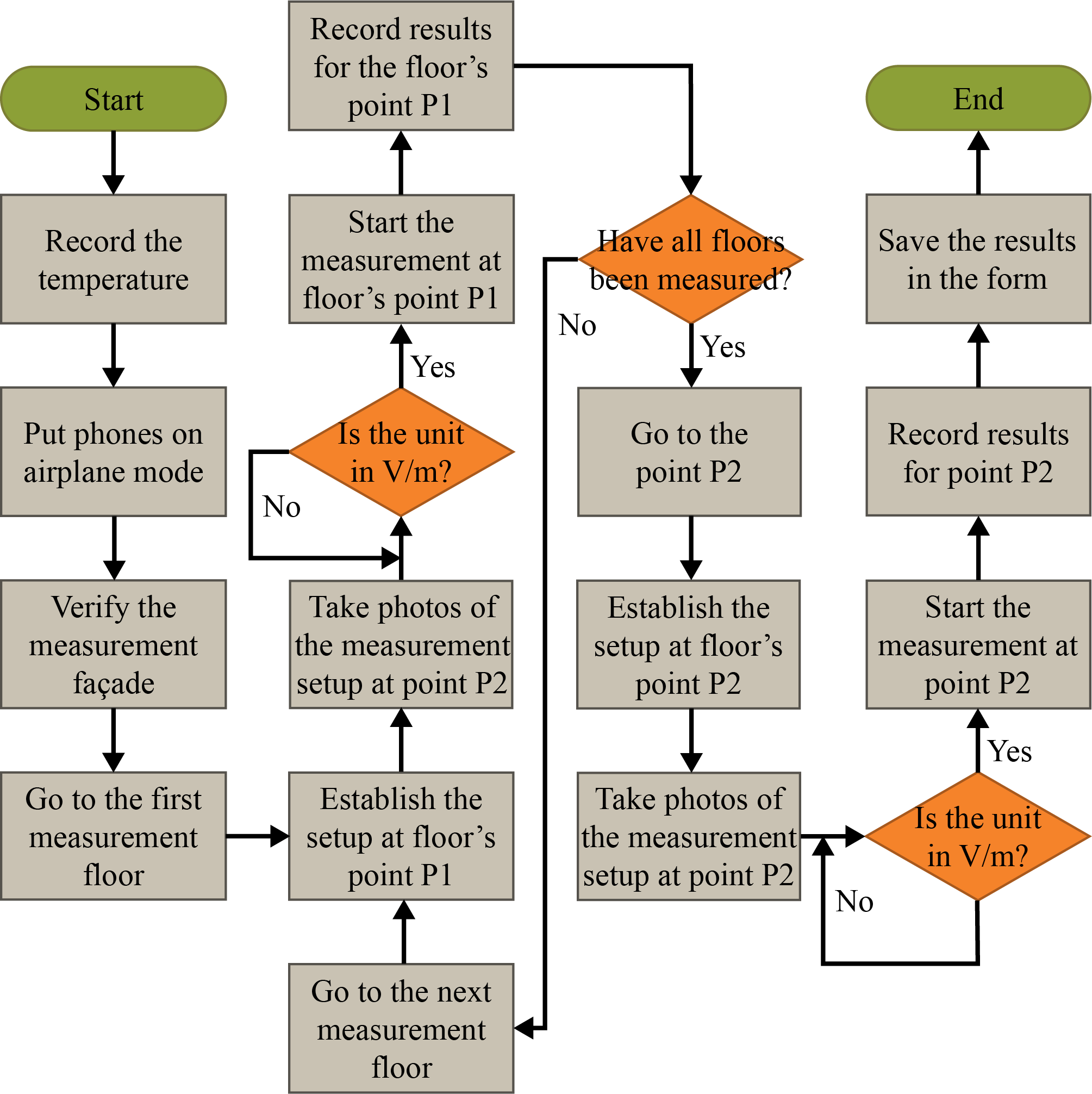}
\caption{Measurement procedure flowchart.}
\label{fig:fluxogramas}
\vspace{-7pt}
\end{figure}

The flowchart in Fig.~\ref{fig:fluxogramas} summarizes the procedure followed in the measurement campaigns described in this section.

\section{Results}
\label{sec:results}

Tab.~\ref{tab:1-minutes-measurement} shows the results of average and peak Electric Field (EF) intensity, and the ERs for the one-minute measurement campaign. 

\begin{table}[H]
    \centering
    \caption{One-minute campaign average, peak, and exposure ratio results.}
    \label{tab:1-minutes-measurement}
	\begin{tabular*}{\columnwidth}{@{\extracolsep\fill}ccccc@{\extracolsep\fill}}
	\hline 
	      \multicolumn{1}{l}{\textbf{\makecell{Measurement \\ Location}}} &
        \multicolumn{1}{l}{\textbf{\makecell{Average EF \\ Intensity \\ (V/m)}}} &
        \multicolumn{1}{l}{\textbf{\makecell{Peak EF \\ Intensity \\ (V/m)}}} &
        \multicolumn{1}{l}{\textbf{\makecell{ER (\%) \\ (778~MHz)}}} &
        \multicolumn{1}{l}{\textbf{\makecell{ER (\%) \\ (88~MHz)}}} \\ 
        \hline
        Street Level & 0.64 & 0.97 & 0.03 & 0.05 \\  
        1\textsuperscript{st} Floor & 0.44 & 0.70 & 0.01 & 0.03 \\  
        2\textsuperscript{nd} Floor & 0.47 & 0.64 & 0.02 & 0.03 \\  
        3\textsuperscript{rd} Floor & 0.50 & 0.73 & 0.02 & 0.03 \\
        4\textsuperscript{th} Floor & 0.79 & 1.03 & 0.04 & 0.08 \\  
        5\textsuperscript{th} Floor & 0.69 & 0.88 & 0.03 & 0.06 \\  
        6\textsuperscript{th} Floor & 1.18 & 1.55 & 0.09 & 0.18 \\  
        7\textsuperscript{th} Floor & 2.30 & 2.75 & 0.36 & 0.69 \\  
        8\textsuperscript{th} Floor & 3.90 & 5.07 & 1.03 & 1.98 \\  
        9\textsuperscript{th} Floor & 7.33 & 9.91 & 3.65 & 7.00 \\  
        10\textsuperscript{th} Floor & 17.16 & 28.47 & 20.02 & 38.38 \\  
        11\textsuperscript{th} Floor & 13.36 & 15.98 & 12.14 & 23.26 \\  
        12\textsuperscript{th} Floor & 6.57 & 8.26 & 2.93 & 5.63 \\  
        13\textsuperscript{th} Floor & 3.12 & 4.06 & 0.66 & 1.27 \\ 
        14\textsuperscript{th} Floor & 1.72 & 2.09 & 0.20 & 0.39 \\  
        15\textsuperscript{th} Floor & 2.22 & 2.90 & 0.34 & 0.64 \\  
        16\textsuperscript{th} Floor & 2.01 & 2.54 & 0.27 & 0.53 \\  
        17\textsuperscript{th} Floor & 1.05 & 1.64 & 0.07 & 0.14 \\  
        18\textsuperscript{th} Floor & 1.06 & 1.30 & 0.08 & 0.15 \\  
        19\textsuperscript{th} Floor & 1.31 & 1.61 & 0.12 & 0.22 \\  
        20\textsuperscript{th} Floor & 0.89 & 1.09 & 0.05 & 0.10 \\ 
        \hline 
	\end{tabular*}
\end{table}

The average electric field intensity ranged from 0.44~V/m on the first level to 17.16~V/m on the 10\textsuperscript{th} floor, which also presented a peak value of 28.47~V/m, the highest measured in this campaign. The average electric field finding of the 10\textsuperscript{th} floor represents 20.02\% of the limit established for the mobile telephony frequency band (38.35~V/m) and 38.38\% of the 27.70~V/m limit established for the broadcast band.

Tab.~\ref{tab:30-minutes-measurement} indicates that the results of the 30-minute measurement campaign on the 9\textsuperscript{th}, 10\textsuperscript{th}, and 11\textsuperscript{th} floors presented a higher average and peak electric field intensity values than the first campaign on these floors. This is caused by the more extended data collection period used in this campaign, which enables a more comprehensive capture of network traffic variations and power emission.

\begin{table}[ht!]
    \centering
    \caption{30-minute campaign average, peak, and exposure ratio results.}
    \label{tab:30-minutes-measurement}
	\begin{tabular*}{\columnwidth}{@{\extracolsep\fill}ccccc@{\extracolsep\fill}}
	\hline 
	      \multicolumn{1}{l}{\textbf{\makecell{Measurement \\ Location}}} &
        \multicolumn{1}{l}{\textbf{\makecell{Average EF \\ Intensity \\ (V/m)}}} &
        \multicolumn{1}{l}{\textbf{\makecell{Peak EF \\ Intensity \\ (V/m)}}} &
        \multicolumn{1}{l}{\textbf{\makecell{ER (\%) \\ (778~MHz)}}} &
        \multicolumn{1}{l}{\textbf{\makecell{ER (\%) \\ (88~MHz)}}} \\ 
        \hline
        Street Level & 0.83 & 1.33 & 0.05 & 0.09 \\  
        9\textsuperscript{th} Floor & 6.74 & 10.18 & 3.09 & 5.92 \\  
        10\textsuperscript{th} Floor & 17.34 & 31.86 & 20.44 & 39.19 \\  
        11\textsuperscript{th} Floor & 13.13 & 18.36 & 11.72 & 22.47 \\
        \hline 
	\end{tabular*}
\end{table}

The 30-minute campaign's average electric field values ranged from 0.83~V/m on the street level to 17.34~V/m on the 10\textsuperscript{th} floor, with the 10\textsuperscript {th} floor also presenting the highest peak value measured (31.86~V/m). The average electric field result of the 10\textsuperscript{th} floor represents 20.44\% of the limit established for the mobile telephony frequency band and 39.19\% of the more restrictive limit for the equipment's frequency range.

\begin{figure*}[!hb]
\centering
\includegraphics[width=0.69\linewidth]{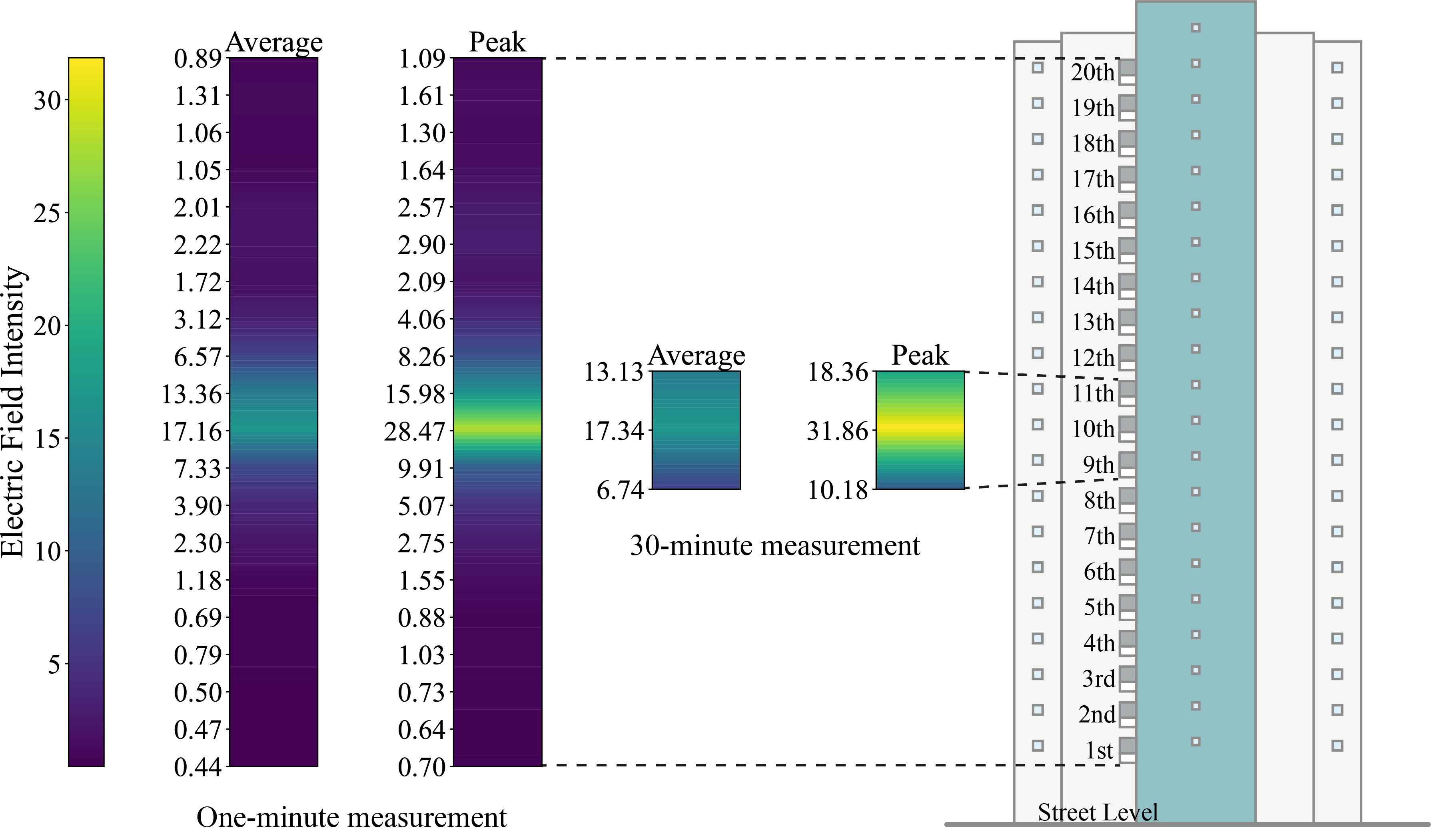}
\caption{Distribution of electric field intensity on floors with the highest recorded values.}
\label{fig:exposition}
\end{figure*}

Fig.~\ref{fig:exposition} illustrates the electric field intensity distribution from both measurement campaigns across the building's floors. The observed behavior aligns with the methodological analysis, as floors distant from the center of the HPBW have lower electric field intensity values. The maximum peak electric field strength was 31.86~V/m, which was 23.95 times higher than the peak value recorded at street level. 

Although the overall trend complies with theoretical estimations, several floors have abnormal values compared to their position. For example, the 5\textsuperscript{th} floor had a higher value than the 6\textsuperscript{th}, while the 18\textsuperscript{th} and 19\textsuperscript{th} floors had higher values than the 17\textsuperscript{th}. This behavior may be explained by the one-minute campaign's measurements being more sensitive to changes introduced by other telecommunications services, such as adjacent Wi-Fi equipment, due to the short data gathering period.

Despite the discrepancy between the floor at the center of the estimated zone of interest and the one determined by measurement, the analysis anticipated the floors with the highest values. Furthermore, these floors showed significantly higher exposure levels than those at ground level, highlighting the importance of taking measurements on floors adjacent to the floor in the center of the vertical zone of interest whenever possible. 

The discrepancy found may result from unpredictable factors, such as differences between the mechanical tilt angle registered in ANATEL’s platform and the angle used in the antenna installation. Another possible contributing factor is the hypothetical presence of an intrinsic electrical downtilt of the antenna model used for transmission. As noted in~\cite{STOTEN_2025}, this parameter cannot be directly determined with the data available on ANATEL’s platform~\cite{anatelsite}. Furthermore, the use of broadband measurement equipment limits the capacity to isolate contributions from each operator or radio access technologies, which could affect the interpretation of the observed disparity.

\section{Conclusions}
\label{sec:conclusions}


This paper presented a case study assessing RF-EMF exposure across every floor of a building located in a critical RF emission scenario, due to its proximity to nearby telecommunications infrastructure operated by multiple operators.

Extending the results of \cite{STOTEN_2025}, our unpublished results showed a significant rise in average and peak electric field intensity values as the measurement floor approached the center of the antenna's Half-power Bandwidth (HPBW). These results emphasize the need for a methodology that considers the configuration of BS antennas when assessing modern RF-EMF exposure scenarios, while also being broadly applicable.

The discrepancy between the target floor and the floor with the highest exposure levels underscores the importance of continuously updating BS parameter databases with reliable information to ensure the accuracy and time efficiency of measurements, such as those reported in this study.

In future works, we plan to conduct measurements using narrowband equipment to understand the contribution of each technology to the total exposure levels in the evaluated scenarios. In addition, we intend to conduct further vertical measurements within buildings, establishing collaboration with condominium management and ANATEL to allow access to apartment interiors when needed.

By establishing partnerships with network operators to obtain the exact antenna model and configuration for the evaluated sites, we aim to improve the accuracy of estimates using the methodology proposed in~\cite{STOTEN_2025}. Based on these measurements, we intend to develop a predictive model of electromagnetic field intensity in buildings exposed to direct BS emissions, accounting for transmission frequency, antenna configuration, and the distance between the antennas and the target building.

Future work will also include measurements to characterize the operational conditions under which different exposure levels are observed, relating electric field intensity to channel conditions, network configuration, and traffic load. In particular, scenarios with co-channel interference, such as dense deployments with frequency reuse, may exhibit elevated exposure levels due to simultaneous transmissions from multiple cells, even when the system operates in interference-limited regimes, highlighting a potential mismatch between measured exposure and effective communication conditions.


\bibliographystyle{IEEEtran}
\bibliography{bibliography}

\newpage

\begin{IEEEbiography}
[{\includegraphics[width=1in,height=1.25in,clip,keepaspectratio]{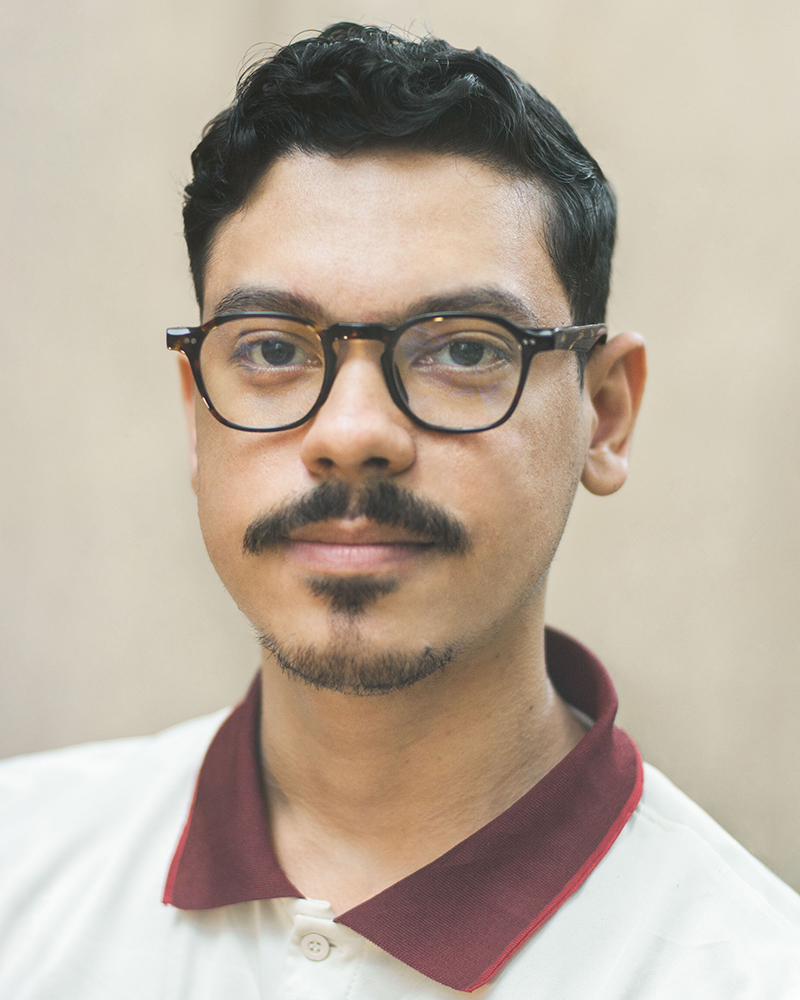}}]
{Ricardo Q. de F. H. Silva} holds a B.S. (2023) and an M.Sc. (2025) in Electrical Engineering from the Federal University of Rio Grande do Norte (UFRN), Brazil. He also earned a technical degree in Electronics from the Federal Institute of Education, Science, and Technology of Rio Grande do Norte (IFRN) in 2017. Currently, he is a Ph.D. candidate and research fellow at the GppCom Research Group, where his work centers on mobile communication systems. Silva has solid experience with network simulation using the ns-3 discrete-event simulator, particularly in Wi-Fi, LTE, and 5G NR scenarios. He also works on OpenRAN deployments leveraging the OpenAirInterface (OAI), open5GS, and srsRAN platforms. Additionally, he contributes to projects focused on RF-EMF exposure assessment, in collaboration with the Brazilian National Telecommunications Agency (ANATEL).
\end{IEEEbiography}

\begin{IEEEbiography}
[{\includegraphics[width=1in,height=1.25in,clip,keepaspectratio]{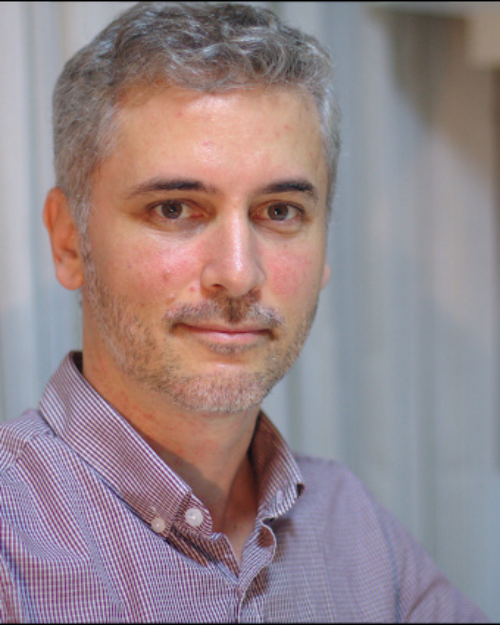}}]
{Marcio E. C. Rodrigues} has a BSc. degree in Telecommunications Engineering from Fluminense Federal University (UFF, Brazil, in 1997) and MSc. (in 2000) and PhD (year 2010) degrees in Electrical
Engineering (Telecommunications) from Pontifical Catholic University of Rio de Janeiro (PUC-Rio,
Brazil), including one year of doctoral stay at ONERA (Toulouse/France, in 2008). In the private
enterprise, he took part in projects and research in the field of radio propagation and wireless com
munications, for mobile operators, ANATEL and Petrobrás, among others. Since 2011, he has been
a professor at UFRN, having a research interest involving radio propagation and wireless channel
modelling, besides Engineering teaching methodologies.
\end{IEEEbiography}

\begin{IEEEbiography}
[{\includegraphics[width=1in,height=1.25in,clip,keepaspectratio]{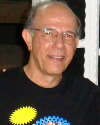}}]
{Fred S. R. Pinheiro} holds a PhD degree in Health Sciences (2017), a master’s and an undergraduate degree in Electrical Engineering (2006, 1975) from UFRN. He holds a specialization degree in Telecommunications Engineering (1996) from UFCG. As an engineer at TELERN and TELEMAR (OI) (1975 to 2002), he was responsible for deploying microwave communication and telephony systems. He was responsible for implementing and managing the communications system that supported Pope John Paul II’s visit to Natal, RN, Brazil, in 1991. He took a specialization course in Rural Telecommunications Planning at the USTTI (United States Government Telecommunications Training Institute), Washington and Los Angeles, 1993. Nowadays, Dr. Fred is a professor and researcher at UFRN.
\end{IEEEbiography}

\begin{IEEEbiography}
[{\includegraphics[width=1in,height=1.25in,clip,keepaspectratio]{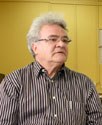}}]
{Gutembergue S. da Silva} holds a PhD in Electrical and Computer Engineering (2015) from UFRN and a master’s degree in Electrical Engineering (1992) from the Federal University of Rio de Janeiro (UFRJ). He also graduated in Electrical Engineering (1975) and Economics (1983) from UFRN. He also has a specialization degree in Telecommunications Engineering (1996) from the Federal University of Campina Grande (UFCG), Brazil. Between 1975 and 2002, he developed activities in planning, design, implementation, and operation of telecommunications systems as an executive at TELERN-TELEBRAS, Brazil. Nowadays, Dr. Gutembergue is a professor and researcher at UFRN, Brazil.
\end{IEEEbiography}

\begin{IEEEbiography}
[{\includegraphics[width=1in,height=1.25in,clip,keepaspectratio]{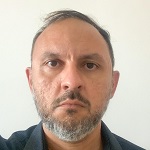}}]
{Halysson B. Mendonça} received his MS degree in mobile communications in 2002. Since 2005, he has been working as an engineer at the National Telecommunications Agency- ANATEL (Agência Nacional de Telecomunicações). He is responsible for the inspection of the technical rules of regulated companies. In 2011, Mr. Mendonça concluded a specialization course in Telecommunications Regulation at INATEL, a multidisciplinary course, including technical, legal, and other regulatory aspects.
\end{IEEEbiography}

\begin{IEEEbiography}
[{\includegraphics[width=1in,height=1.25in,clip,keepaspectratio]{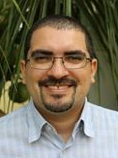}}]
{Vicente A. de Sousa Jr.} received his B.S., M.S and Ph.D.  degrees in Electrical Engineering from the Federal University of Ceará (UFC), Fortaleza, CE, Brazil, in 2001,  2002  and  2009,  respectively.  Between 2001 and 2006, he developed solutions to UMTS/WLAN interworking for   UFC   and   Ericsson of   Brazil. Between 2006 and 2010, he contributed to WIMAX standardization and Nokia’s product as a researcher at the Institute of Technological Development (INdT). Dr.  Sousa is now a professor and the head of the GppCom Research Group at UFRN. He contributed to 5G open RAN projects supported by Lenovo and to NIR measurements and is working with evaluation projects supported by ANATEL Agency.
\end{IEEEbiography}

\end{document}